\documentstyle[12pt]{article}
\input style
\def\baselinestretch{1.65}
\catcode`\@=11
\def\section{\@startsection{section}{1}{0.0ex}
                   {3.5ex plus -1.0ex minus -0.2ex}
                   {2.3ex plus 0.2ex}{\bf}}
\def\subsection{\@startsection{subsection}{2}{0.0ex}
                        {3.25ex plus 1ex minus .2ex}
                        {1.5ex plus .2ex}{\bf}}
\catcode`@=12
\input math_macros
\begin{document}
\def\thefootnote{\fnsymbol{footnote}}
\FERMILABPub{94/061--T}
\begin{titlepage}
\vspace*{-0.7in}
\begin{flushright}
        FERMILAB--PUB--94/061--T\\
        OSU Preprint 286\\
        March 1994\\
\end{flushright}
\vspace{-0.40in}
\begin{center}
{\large \bf Construction of Fermion Mass Matrices\\
	Yielding Two Popular Neutrino Scenarios}\\
%\vfill
\vskip 0.20in
        {\bf Carl H. ALBRIGHT}\\
 Department of Physics, Northern Illinois University, DeKalb, Illinois
60115\footnote{Permanent address}\\[-0.2cm]
        and\\[-0.2cm]
 Fermi National Accelerator Laboratory, P.O. Box 500, Batavia, Illinois
60510\footnote{Electronic address: ALBRIGHT@FNALV}\\
        and\\
        {\bf Satyanarayan NANDI}\\
 Department of Physics, Oklahoma State University, Stillwater, Oklahoma
        74078\footnote{Electronic address: PHYSSNA@OSUCC}\\
\end{center}
%\vskip 0.5in
\vfill
\begin{abstract}
A new procedure proposed recently enables one to start from the quark and
lepton mass and mixing data at the low scale and construct mass matrices
which exhibit simple SO(10) structure at the SUSY GUT scale.  We elaborate
here on the numerical details which led us to an SO(10) model for the quark
and lepton mass matrices that explain the known quark data at the low scale
along with the observed depletions of solar- and atmospheric-neutrinos.  We
also apply the procedure to a second scenario incorporating the solar-neutrino
depletion and a 7 eV tau-neutrino for the cocktail model of mixed dark matter
but find the SO(10) model deduced in this case does not exhibit as simple a
structure as that observed for the first scenario.
\end{abstract}
\end{titlepage}
\section{GENERAL APPROACH AND APPLICATION}

In a recent letter,$^1$ the authors sketched a new approach which
enables one to make use of quark and lepton mass and mixing data at the low
energy scale to construct an SO(10)-symmetric fermion mass matrix model at the
supersymmetric grand unified scale from which the low energy results can be
derived.  This ``bottom-up approach'' should be contrasted with the usual
procedure, where one introduces some ansatz for the fermion mass matrices at
the
grand unification scale, from which certain predictions can be made at the
low scale.  In proposing such an ansatz, one must take care not to violate
any of the known data at the low scale.  This conventional method has a very
extensive literature dating from the early work of Fritzsch$^2$ up to the
present
models proposed by many authors$^3$ in SO(10) SUSY GUTS, where the evolution
of the Yukawa couplings from the SUSY GUT scale to the weak scale plays a
major role.  In some of the most recent work along these lines, authors have
attempted to
impose$^4$ as many texture zeros as possible (typically five or six) for the up
and down quark matrices, or considered$^5$ just one $\bf{10}$ and one
$\bf{126}$
Higgs representations of SO(10) contributing to the mass matrices, and then
deduced the consequences of these assumptions.  In doing so, they find that
the combined tau-neutrino mass and mixing angles do not accurately fit
either the popular cocktail model$^6$ of mixed dark matter or the oscillation
explanation of the observed atmospheric depletion$^7$ of muon-neutrinos.

In our approach, on the other hand, one must input all the known or presumed
known
masses and mixings in order to construct numerically the mass matrices by a
method proposed in the quark context by Kusenko.$^8$  Since the neutrino mass
and
mixing data are not well known at this time, many scenarios can be considered
for the starting point.  In the letter cited above, we illustrated the
procedure with neutrino data extracted from the non-adiabatic
Mikheyev-Smirnov-Wolfenstein$^9$ (MSW)
interpretation of the observed solar electron-neutrino depletion$^{10}$ and
from
the muon-neutrino and tau-neutrino mixing interpretation of the observed
atmospheric muon-neutrino depletion effect.$^7$   Here we shall elaborate on
the numerical details which led to the mass matrices in the SO(10) framework
proposed in Ref. 1 and apply the same technique to a second neutrino
scenario involving the same solar electron-neutrino depletion, but now
in the presence of a 7 eV tau-neutrino which provides 30\% of the missing
dark matter, the rest arising from the supersymmetric neutralinos in the
cocktail model$^6$ of mixed dark matter.  In this new approach, we can identify
what assumptions regarding texture zeros and minimal Higgs content must be
relaxed in order to obtain accurate model fits to the two scenarios in
question.

We first restate here the basic steps for the construction of the quark and
lepton mass matrices in the new approach:
\begin{itemize}
\item   Start from the known and/or presumed-known quark and lepton masses,
        $m_q$'s, $m_{\ell}$'s and $m_{\nu}$'s; and quark and lepton mixing
        matrices, $V_{CKM}$ and $V_{LEPT}$, at the low scales.
\item   Evolve the masses and mixing matrices to the SUSY GUT scale
        using the appropriate renormalization group equations
        (RGEs) for the minimal supersymmetric standard model (MSSM).
\item   Construct complex symmetric $M^U$, $M^D$, $M^E$, and $M^{N_{eff}}$
        matrices for the up and down quarks, charged leptons and
        light neutrinos using a modified procedure of Kusenko$^8$ described
        later.  Two parameters $x_q$ and $x_{\ell}$ allow one to adjust the
        diagonal/off-diagonal nature of the quark and lepton mass matrices.
\item   Vary $x_q$ and $x_{\ell}$ systematically over their support regions
        while searching for as many pure ${\bf 10}$ or pure ${\bf 126}$
        $SO(10)$ contributions to the matrix elements as possible.
\item   For the ``best'' choice of $x_q$ and $x_{\ell}$, construct a simple
        model of the mass matrices with as many texture zeros as possible.
\item   Evolve the mass eigenvalues and mixing matrices determined from the
        model at the SUSY GUT scale to the low scale and compare the results
        with the starting input data.
\end{itemize}
In Sect.~II we shall assign values to the quark and lepton masses and
mixings for the two scenarios investigated in this paper.  Evolution to the
SUSY GUT scale is discussed in Sect.~III.  Numerical construction of the mass
matrices is explained in Sect.~IV, followed by the SO(10) model constructions
in Sect.~V.  There we also compute the quark and lepton mass eigenvalues and
mixing matrices in the two models and evolve the results downward to the
low scales to compare the SO(10) model results with the original quark and
lepton input parameters.  In Sect.~VI we draw our conclusions.

\section{MASSES AND MIXINGS AT THE LOW SCALES}

Uncertainties in the quark masses and mixings lie within relatively broad
bounds.  The light quark mass ratios are quite accurately determined from
current algebra, while the absolute values are more uncertain.$^{11}$  This is
especially true for the strange quark mass.  We shall adopt as input the
central values at 1 GeV quoted by Gasser and Leutwyler$^{12}$ about ten years
ago,
which are as good as any obtained since then.  The $c$ and $b$ quark masses are
specified at their running mass scales, while the corresponding top quark mass
is much less certain$^{13}$ since its discovery has yet to be made.
With $m_t^{phys} \sim 160$ GeV, we adopt as starting input the following
quark masses$^{12}$
$$\begin{array}{rlrl}
        m_u(1 {\rm GeV})&= 5.1\ {\rm MeV},& \qquad m_d(1 {\rm GeV})&= 8.9\
                {\rm MeV}\nonumber\\
        m_c(m_c)&= 1.27\ {\rm GeV},& \qquad m_s(1 {\rm GeV})&= 175\ {\rm MeV}
                \cr
        m_t(m_t)&= 150\ {\rm GeV},& \qquad m_b(m_b)&\simeq 4.25\ {\rm GeV}\cr
  \end{array}\eqno(2.1a)$$
The Cabibbo-Kobayashi-Maskawa (CKM) mixing matrix,$^{14}$ on the other hand, is
becoming better known with time, its main uncertainties$^{15}$ being $V_{cb}$,
and especially the $|V_{ub}/V_{cb}|$ ratio and the CP-violating phase.
We adopt at the weak scale the central values
$$V_{CKM} = \left(\matrix{0.9753 & 0.2210 & (-0.283 -0.126i)\times 10^{-2}\cr
                -0.2206 & 0.9744 & 0.0430\cr
                0.0112 -0.0012i & -0.0412 -0.0003i & 0.9991\cr}\right)
        \eqno(2.1b)$$
where we have assumed a value of 0.043 for $V_{cb}$ and applied strict
unitarity to determine $V_{ub},\ V_{td}$ and $V_{ts}$.

In contrast, although the charged lepton masses are precisely known, the
lepton mixings and neutrino masses remain uncertain.  But a reasonable starting
point has emerged with the increased knowledge gained from the solar
neutrino experiments$^{10}$ involving the chlorine experiments of Davis, the
water
Cerenkov experiments of the Irvine-Michigan-Brookhaven and the Kamiokande
collaborations and the recent gallium experiments of the SAGE and GALLEX
collaborations.  Taken together, the depleted electron-neutrino fluxes
observed compared with the standard solar model predictions suggest that
non-adiabatic MSW resonant conversion$^9$ of the electron-neutrinos into
muon-neutrinos
in the solar interior is most likely responsible.  The central values deduced
for this effect are $\delta m_{12}^2 \sim 5 \times 10^{-6}\ {\rm eV^2}$ and
$\sin^2 2\theta_{12} \sim 8 \times 10^{-3}$.

With regard to the tau-neutrino mass and mixings, two scenarios
are popular and conflicting, if one does not assume near degeneracy of the
neutrino masses or the existence of a new sterile neutrino.  We shall make no
such assumption.  In the first scenario
labeled (A), one suggests that muon-neutrinos
oscillate into tau-neutrinos on their passage through the atmosphere and hence
deplete the flux of muon-neutrinos relative to electron-neutrinos to explain
the observed atmospheric depletion.$^{16}$  The central values
for this interpretion are: $\delta m_{23}^2 \sim 2 \times 10^{-2}\ {\rm eV^2}$
and $\sin^2 2\theta_{23}~\sim 0.5$.

Alternatively, in the second scenario labeled (B) and popularly known as
the cocktail model,$^6$ one speculates that
tau-neutrinos account for 30\% of the dark matter as a source of missing hot
dark matter, while supersymmetric neutralinos serve as a source of cold dark
matter and account for the remaining 70\%.  For this case, the simplest
interpretation is that the tau-neutrino
has a mass of 7 eV.  The present accelerator data on $\nu_{\mu}\ -\ \nu_{\tau}$
oscillations then place an upper limit of $\sin^2 2\theta_{23}\ \ltap\ 10^{-3}$
on the mixing angle.$^{17}$

We take for the lepton input in neutrino scenario (A)
$$\begin{array}{rlrl}
        m_{\nu_e}&= 0.5 \times 10^{-6}\ {\rm eV},& \qquad m_e&= 0.511\ {\rm
                MeV}\nonumber\\
        m_{\nu_{\mu}}&= 0.224 \times 10^{-2}\ {\rm eV},& \qquad m_{\mu}&=
                105.3\ {\rm MeV}\cr
        m_{\nu_{\tau}}&= 0.141\ {\rm eV},& \qquad m_{\tau}&= 1.777\ {\rm
                GeV}\cr \end{array}\eqno(2.2a)$$
and
$$V^{(A)}_{LEPT} = \left(\matrix{0.9990 & 0.0447 & (-0.690 -0.310i)
		\times 10^{-2}\cr -0.0381 -0.0010i & 0.9233 & 0.3821\cr
                0.0223 -0.0030i & -0.3814 & 0.9241\cr}\right) \eqno(2.2b)$$
We have simply assumed a value of for the electron-neutrino mass to which our
analysis is not very sensitive and constructed
the lepton mixing matrix$^{18}$ by making use of the unitarity conditions with
the same phase in (2.1b) and (2.2b).  For scenario (B), we use
$$\begin{array}{rlrl}
        m_{\nu_e}&= 0.5 \times 10^{-6}\ {\rm eV},& \qquad m_e&= 0.511\ {\rm
                MeV}\nonumber\\
        m_{\nu_{\mu}}&= 0.224 \times 10^{-2}\ {\rm eV},& \qquad m_{\mu}&=
                105.3\ {\rm MeV}\cr
        m_{\nu_{\tau}}&= 7.0\ {\rm eV},& \qquad m_{\tau}&= 1.777\ {\rm
                GeV}\cr \end{array}\eqno(2.3a)$$
and
$$V^{(B)}_{LEPT} = \left(\matrix{0.9990 & 0.0447 & (-0.289 -0.129i)
		\times 10^{-2}\cr  -0.0446 & 0.9989 & 0.0158\cr
                0.0036 -0.0013i & -0.0157 -0.0001i & 0.9998\cr}\right)
		\eqno(2.3b)$$
In scenario (A) the tau-neutrino mass is of the order of 0.1 eV, while
the 23 element of the leptonic mixing matrix is large, while in scenario (B)
the tau-neutrino mass is fifty times larger, but the 23 element of the
leptonic mixing matrix is very small; in fact, this mixing matrix is very
close to the identity.

\section{EVOLUTION TO THE SUSY GUT SCALE}

We now evolve the low energy data to the SUSY GUT scale, where any simplicity
due to the SO(10) symmetry should apply.  In order to use analytic expressions
for the running variables, we shall use the one-loop renormalization group
equations (RGEs) with numbers taken from the work of Naculich.$^{19}$  The
supersymmetry breaking scale is assumed to lie at $\mu_{SUSY} = 170$ GeV, while
the GUT scale, where the gauge couplings are unified, occurs at
$\bar{\mu} = 1.2 \times 10^{16}$ GeV.

The connection between the running mass $m_{\alpha}$ of a fermion and its
corresponding Yukawa coupling $y_{\alpha}$ is defined by
$$m_{\alpha} = y_{\alpha}(v/\surd{2})\left\{\matrix{\sin \beta, \quad
	\alpha = u, c, t, \nu_e, \nu_{\mu}, \nu_{\tau}\cr
		\cos \beta, \quad \alpha = d, s, b, e, \mu, \tau\ \ \cr}\right.
	\eqno(3.1a)$$
where $\tan \beta = v_u/v_d$ is the ratio of the up quark to the down quark
VEVs and $v = 246$ GeV, the electroweak symmetry-breaking scale.
The Yukawa coupling running between the 1 GeV scale for the light quarks and
leptons or the running mass scale for the heavy quarks and the supersymmetry
breaking scale $\mu_{SUSY}$ is governed by the gauge couplings and can be
summarized in terms of ratios $\eta_{\alpha}$ of the couplings at the low
scale to those at $\mu_{SUSY}$.  We shall use
$$\begin{array}{rl}
	\eta_u&= \eta_d = 2.17, \quad \eta_s = 2.16 \nonumber\\
        \eta_c&= 1.89, \quad \eta_t = 1.00, \quad \eta_b = 1.47 \cr
	\eta_{\nu_e}&= \eta_{\nu_{\mu}} = \eta_{\nu_{\tau}} = 1.03 \cr
	\eta_e&= \eta_{\mu} = 1.03, \quad \eta_{\tau} = 1.02 \cr
\end{array}\eqno(3.1b)$$

Only the third family quark and charged lepton Yukawa couplings are
assumed$^{19}$ to contribute
to the nonlinear part of the Yukawa coupling evolution from the supersymmetry
breaking scale to the GUT scale.  In this approximation one finds that the
$\mu_{SUSY}$ scale couplings are given in terms of the GUT scale Yukawa
couplings by
$$\begin{array}{rlrl}
y_u(\mu_{SUSY})&= \bar{y}_u A_u B^3_t,& \quad y_d(\mu_{SUSY})&=
	\bar{y}_d A_d B^3_b B_{\tau}\nonumber\\
y_c(\mu_{SUSY})&= \bar{y}_c A_u B^3_t,& \quad y_s(\mu_{SUSY})&=
	\bar{y}_s A_d B^3_b B_{\tau}\cr
y_t(\mu_{SUSY})&= \bar{y}_t A_u B^6_t B_b,& \quad y_b(\mu_{SUSY})&=
	\bar{y}_b A_d B^6_b B_{\tau}B_t\cr
y_{\nu_e}(\mu_{SUSY})&= \bar{y}_{\nu_e} A_n,& \quad y_e(\mu_{SUSY})&=
	\bar{y}_e A_e B^3_b B_{\tau}\cr
y_{\nu_{\mu}}(\mu_{SUSY})&= \bar{y}_{\nu_{\mu}} A_n,& \quad y_{\mu}
	(\mu_{SUSY})&= \bar{y}_{\mu} A_e B^3_b B_{\tau}\cr
y_{\nu_{\tau}}(\mu_{SUSY})&= \bar{y}_{\nu_{\tau}} A_n,& \quad y_{\tau}
	(\mu_{SUSY})&= \bar{y}_{\tau} A_e B^3_b B^4_{\tau}\cr
\end{array}\eqno(3.2a)$$
where the gauge evolution factors $A_{\alpha}$ are equal to
$$\begin{array}{rlrl}
A_u&= 3.21,& \quad A_d&= 3.13 \nonumber\\
A_n&= 1.37,& \quad A_e&= 1.48 \cr
\end{array}\eqno(3.2b)$$
and the Yukawa evolution factors are approximately equal to
$$\begin{array}{rlrl}
B_t \simeq&\left[1 + \bar{y}^2_t K_u\right]^{-1/12},&\qquad K_u = 8.65
	\nonumber\\
B_b \simeq&\left[1 + \bar{y}^2_b K_d\right]^{-1/12},&\qquad K_d = 8.33\cr
B_{\tau} \simeq&\left[1 + \bar{y}^2_{\tau} K_e\right]^{-1/12},&\qquad K_e =
	3.77\cr \end{array}\eqno(3.2c)$$

By combining (3.1) and (3.2), we can find the Yukawa couplings at the grand
unification scale.  In doing so, we adjust $m_b(m_b)$ and $\tan \beta$
so that complete Yukawa unification$^{20}$ is achieved at $\bar{\mu}$, i.e.,
$\bar{m}_{\tau} = \bar{m}_b = \bar{m}_t/\tan \beta$.  This is accomplished by
choosing $m_b(m_b) = 4.09$ GeV at the running $b$ quark mass scale$^{21}$ and
$\tan \beta = 48.9$.  The evolved masses for the quarks at $\bar{\mu}$ are
then found to be
$$\begin{array}{rlrl}
        \bar{m}_u&= 1.098\ {\rm MeV},& \qquad \bar{m}_d&= 2.127\
                {\rm MeV}\nonumber\\
        \bar{m}_c&= 0.314\ {\rm GeV},& \qquad \bar{m}_s&= 42.02\ {\rm MeV}\cr
        \bar{m}_t&= 120.3\ {\rm GeV},& \qquad \bar{m}_b&= 2.464\ {\rm GeV}\cr
	\end{array}\eqno(3.3a)$$
and for the leptons in scenario (A)
$$\begin{array}{rlrl}
        \bar{m}_{\nu_e}&= 0.581 \times 10^{-6}\ {\rm eV},& \qquad \bar{m}_e&=
                0.543\ {\rm MeV}\nonumber\\
        \bar{m}_{\nu_{\mu}}&= 0.260 \times 10^{-2}\ {\rm eV},& \qquad
                \bar{m}_{\mu}&= 111.9\ {\rm MeV}\cr
        \bar{m}_{\nu_{\tau}}&= 0.164\ {\rm eV},& \qquad \bar{m}_{\tau}&=
                2.464\ {\rm GeV}\cr \end{array}\eqno(3.3b)$$
Since the third family terms control the Yukawa couplings in the RGEs, only
the following $V_{CKM}$ and $V_{LEPT}$ mixing matrix elements evolve in leading
order and result in
$$\begin{array}{rlrl}
        \bar{V}_{ub}&= (-0.2163 -0.0963i) \times 10^{-2},&\qquad
                \bar{V}_{13}&= (-0.634 -0.285i) \times 10^{-2}\nonumber\\
        \bar{V}_{cb}&= 0.0329&\qquad \bar{V}_{23}&= 0.3508\cr
        \bar{V}_{td}&= 0.0086 -0.0009i&\qquad \bar{V}_{31}&= 0.0205 -0.0028i\cr
        \bar{V}_{ts}&= -0.0315 -0.0002i&\qquad \bar{V}_{32}&= -0.3502\cr
                \end{array}\eqno(3.3c)$$
while the other mixing matrix elements receive smaller corrections
which can be neglected; however, in doing so the unitarity of the mixing
matrices is not quite preserved.

In scenario (B), $\bar{m}_{\nu_{\tau}}$ should be replaced by 8.127\ {\rm eV},
and the lepton mixing matrix elements in (3.3c) by
$$\begin{array}{rlrl}
        \bar{V}_{13}&= (-0.265 -0.118i) \times 10^{-2}
        	&\qquad \bar{V}_{23}&= 0.0145\nonumber\\
        \bar{V}_{31}&= 0.0033 -0.0012i &\qquad \bar{V}_{32}&= -0.0144 -0.0001i
	   	\cr \end{array}\eqno(3.3d)$$

\section{NUMERICAL CONSTRUCTION OF MASS MATRICES}

Having found the masses and mixing matrices at the GUT scale, we can now
construct numerically the quark and lepton mass matrices by making use of a
procedure suggested by Kusenko$^8$ for the quark mass matrices.  Since the
quark mixing matrix $V_{CKM}$ of the charged-current couplings in
the mass bases is unitary and represents an element of the unitary group U(3),
one can express it in terms of one Hermitian generator of the
corresponding U(3) Lie algebra times a phase parameter $\alpha$ by writing
	$$ V_{CKM} = U'_L U^{\dagger}_L = \exp(i\alpha H) \eqno(4.1a)$$
where
	$$i\alpha H = \sum^3_{k=1}(\log v_k){{\prod_{i\neq k}(V_{CKM}
		- v_i I)}\over{\prod_{j\neq k}(v_k - v_j)}} \eqno(4.1b)$$
in terms of the eigenvalues $v_j$ of $V_{CKM}$ by making use of Sylvester's
theorem.$^{22}$  The transformation matrices from the weak to the mass bases
are
given in terms of the same generator but modified phase parameters such that
$$U'_L = \exp(i\alpha Hx_q),\qquad U_L = \exp\left[i\alpha H(x_q - 1)\right]
	\eqno(4.2)$$
and relation (4.1a) is preserved.

The quark mass matrices in the weak basis are then related to those in the
diagonal mass basis by
$$M^U = U'^{\dagger}_L D^U U'_R,\qquad M^D = U^{\dagger}_L D^D U_R\eqno(4.3a)$$
where $D^U$ and $D^D$ are the diagonal matrices in the mass bases with entries
taken from (3.3a).  In what follows, we shall be interested in constructing
quark and lepton mass matrices in the SO(10) framework which are complex
symmetric.  This requires that only the $\bf{10}$ and $\bf{126}$
irreducible
representations of SO(10) develop vacuum expectation values, while the
antisymmetric $\bf{120}$ does not.  In the higher grand unified groups
exhibiting family symmetry such as SO(14) or SO(18), the complex symmetric
representations are naturally selected.
If we impose this restriction on (4.3a), we can eliminate the
transformation matrices for the right-handed fields in favor of
$$M^U = U'^{\dagger}_L D^U U'^{\dagger T}_L,\qquad M^D = U^{\dagger}_L D^D
        U^{\dagger T}_L\eqno(4.3b)$$
The parameter $x_q$ then controls the diagonal/off-diagonal nature of the
mass matrices, where the up quark mass matrix is diagonal for $x_q = 0$, while
the down quark mass matrix is diagonal for $x_q = 1$.  It suffices
to expand $V_{CKM},\ U'_L$ and $U_L$ to third order in $\alpha$ in order to
obtain accurate expressions for the mass matrices $M^U$ and $M^D$.

A similar argument can be applied in order to construct the light neutrino and
charged lepton mass matrices, $M^{N_{eff}}$ and $M^E$, from the lepton masses
and $V_{LEPT}$ mixing
matrix.  Here the generator and phase parameter are different as $V_{CKM}$ and
its eigenvalues $v_i$ are replaced by $V_{LEPT}$ and $v'_i$, but relations
similar to (4.1) through (4.3) still obtain with $x_{\ell}$ replacing $x_q$.

In order to complete the construction of the mass matrices, we must select
pairs of values for $x_q$ and $x_{\ell}$ lying in the unit square support
region, $0 \leq x_q, x_{\ell} \leq 1$.  For this purpose, we search for
a simple SO(10) structure for the mass matrices.  In the SO(10) framework,
the renormalizable Yukawa interaction Lagrangian for the non-supersymmetric
fermions is given by
$${\cal L}_Y = -\sum_i{\bar{\psi^c}}^{(16)}f^{(10_i)}\psi^{(16)}
		\phi^{(10_i)} -\sum_j{\bar{\psi^c}}^{(16)}f^{(126_j)}
		\psi^{(16)}{\bar{\phi}}^{(126_j)} + {\rm h.c.}\eqno(4.4a)$$
where the $f$'s represent Yukawa coupling matrices.   For fermions in the
fundamental ${\bf 16}$ representation, the only Higgs fields allowed lie in
the symmetric ${\bf 10}$ and ${\bf 126}$ representations and the antisymmetric
${\bf 120}$.  As is customary, we ignore the latter, so the mass matrices are
complex symmetric and given by
$$\begin{array}{rl}
        M^U&= \sum_i f^{(10_i)}v_{ui} + \sum_j f^{(126_j)}w_{uj}\nonumber\\
        M^D&= \sum_i f^{(10_i)}v_{di} + \sum_j f^{(126_j)}w_{dj}\cr
        M^{N_{Dirac}}&= \sum_i f^{(10_i)}v_{ui} - 3\sum_j f^{(126_j)}w_{uj}\cr
        M^E&= \sum_i f^{(10_i)}v_{di} - 3\sum_j f^{(126_j)}w_{dj}\cr
                \end{array}\eqno(4.4b)$$
where $v_{ui}$ and $w_{uj}$ are the ${\bf 10}$ and ${\bf 126}$ VEV
contributions to the up quark and Dirac neutrino matrices, and similarly
for the down quark and charged lepton contributions.  The equations in (4.4b)
can be inverted to determine the sum of the ${\bf 10}$ and sum of the ${\bf
126}$ contributions separately.  At this stage we do not know how many
${\bf 10}$ and ${\bf 126}$ representations of each type are necessary.

By varying the $x_q$ and $x_{\ell}$ parameters over the unit square support
region and by allowing all possible signs to appear in the diagonal matrix
entries of $D^U,\ D^D,\ D^E$ and $D^{N_{eff}}$, we can search for a set of
mass matrices which have either pure ${\bf 10}$ or pure ${\bf 126}$ structure
for as many matrix elements as possible.
Such a preferred choice is found for scenario (A) with $x_q = 0$ and $x_{\ell}
= 0.88$ for which the mass matrices are constructed to be
$$M^U = diag (-.1098\times 10^{-2},\ 0.3140,\ 120.3) \eqno(4.5a)$$\\[-0.3in]
$$\begin{array}{rl}
M^D&= \left(\matrix{(-.8847 + .1072i)\times 10^{-4}& (-.9688 - .0080i)\times
	10^{-2}& (-.4967 - .2371i)\times 10^{-2}\cr
	(-.9688 - .0080i)\times 10^{-2}& -.3705\times 10^{-1}&
	(.8221 + .0001i)\times 10^{-1}\cr
	(-.4967 - .2371i)\times 10^{-2}& (.8221 + .0001i)\times 10^{-1}&
	2.460\cr}\right)\cr \end{array}\eqno(4.5b)$$\\[-0.3in]
$$\begin{array}{rl}
M^E&= \left(\matrix{(-.5339 + .0027i)\times 10^{-3}& (.4135 - .0425i)\times
	10^{-3}& (-.4005 - .0837i)\times 10^{-2}\cr
	(.4135 - .0425i)\times 10^{-3}& 0.1160& 0.1020\cr
	(-.4005 - .0837i)\times 10^{-2}& 0.1020& 2.453\cr}\right) \cr
\end{array}\eqno(4.5c)$$
in units of GeV and
$$\begin{array}{rl}
M^{N_{eff}} = \left(\matrix{(.4839 + .1534i)\times 10^{-4}&
	(-.9059 - .1304i)\times 10^{-3}& (.3023 + .0374i)\times 10^{-2}\cr
	(-.9059 - .1304i)\times 10^{-3}& (.1465 - .0001i)\times 10^{-1}&
	(-.5065 + .0002i)\times 10^{-1}\cr
	(.3023 + .0374i)\times 10^{-2}& (-.5065 + .0002i)\times 10^{-1}&
	0.1502\cr}\right)\cr\end{array}\eqno(4.5d)$$
in units of electron volts.

For scenario (B), the simplest SO(10) construct, with as many pure ${\bf 10}$
or pure ${\bf 126}$ matrix elements as possible, is obtained with $x_q = 0.5$
and $x_{\ell} = 0$, but further investigation reveals there is only one
texture zero.  In place of that, we examine a slightly more complicated
choice which parallels that of scenario (A) with four texture zeros for
which $x_q = 0$ and $x_{\ell} = 0.3$.  The numerical matrices in this case
are exactly the same for the quarks as given in (4.5a,b), while the lepton
matrices are replaced by
$$\begin{array}{rl}
M^E&= \left(\matrix{(-.4256 + .0079i)\times 10^{-3}& (.3469 - .0021i)\times
	10^{-2}& (-.4781 - .2042i)\times 10^{-2}\cr
	(.3469 - .0021i)\times 10^{-2}& 0.1120& 0.2386\times 10^{-1}\cr
	(-.4781 - .2042i)\times 10^{-2}& 0.2386\times 10^{-1}& 2.463\cr}
	\right) \cr \end{array}\eqno(4.6a)$$
in units of GeV and
$$\begin{array}{rl}
M^{N_{eff}} = \left(\matrix{(.5994 + .5325i)\times 10^{-5}&
	 (.0240 - .1243i)\times 10^{-4}& (.7495 + .2887i)\times 10^{-2}\cr
	 (.0240 - .1243i)\times 10^{-4}& - .2448\times 10^{-2}&
	 (-.3518 + .0008i)\times 10^{-1}\cr
	(.7495 + .2887i)\times 10^{-2}& (-.3518 + .0008i)\times 10^{-1}&
	8.127\cr}\right)\cr\end{array}\eqno(4.6b)$$
in units of electron volts.

\section{IDENTIFICATION OF SO(10) MODELS AND PREDICTIONS OF THE MODELS}

\subsection{Neutrino Scenario (A) with Atmospheric Neutrino Depletion}

In order to construct an SO(10) model which closely approximates the numerical
matrices found in (4.5), we take note of the following features.
The up quark matrix $M^U$ is diagonal and the structure for $M^D$ and
$M^E$ is approximately given by
$$M^D \sim M^E \sim \left(\matrix{10,126 & 10,126 & 10\cr 10,126 & 126 & 10\cr
        10 & 10 & 10\cr}\right)\eqno(5.1a)$$
as observed from Eqs. (4.4b),
with $M^D_{11},\ M^E_{12}$ and $M^E_{21}$ anomalously small, i.e., smaller than
expected when compared to the pattern for the other elements.  We shall, in
fact, assume that these elements exhibit texture zeros.$^4$  Most of the
other elements are essentially real with the 13 and 31 elements of $M^D$ and
$M^E$ the major exceptions.  Hence we let only the latter elements be
complex.$^{23}$
If we also assume that the same ${\bf 10}$ and ${\bf 126}$ VEVs contribute,
respectively, to the 33 and 22 diagonal elements of $M^U$ and $M^D$, we find
$$M^U \sim M^{N_{Dirac}} \sim diag(10,126;\ 126;\ 10)\eqno(5.1b)$$

If we now seek as simple a structure as possible for the four matrices,
we are led numerically to the following choices for the Yukawa coupling
matrices at the GUT scale
$$\begin{array}{rlrl}
f^{(10)}&= diag(0,\ 0,\ f^{(10)}_{33}),&\qquad f^{(126)}&= diag(f^{(126)}_{11},
        \ f^{(126)}_{22},\ 0)\nonumber\\[0.1in]
f^{(10')}&= \left(\matrix{f^{(10')}_{11}&f^{(10')}_{12}&f^{(10')}_{13}\cr
        f^{(10')}_{12}& 0 &f^{(10')}_{23}\cr
        f^{(10')}_{13}&f^{(10')}_{23}& 0\cr}\right),&\qquad
f^{(126')}&= \left(\matrix{0 & f^{(126')}_{12}& 0\cr
        f^{(126')}_{12}& 0 & 0\cr 0 & 0 & 0\cr}\right)\end{array}\eqno(5.2)$$
The model requires a minimum of two ${\bf 10}$'s and two ${\bf 126}$'s of
SO(10) with ${\bf 10'}$ and ${\bf 126'}$ having no VEVs in the up direction.
There are four texture zeros in the $M^U$ and $M^D$ matrices taken together.
The four mass matrices are then given by
$$\eqalignno{
        M^U&= f^{(10)}v_u + f^{(126)}w_u &(5.3a)\cr
        M^{N_{Dirac}}&= f^{(10)}v_u - 3 f^{(126)}w_u &(5.3b)\cr
        M^D&= f^{(10)}v_d + f^{(126)}w_d + f^{(10')}v'_d + f^{(126')}w'_d
                &(5.3c)\cr
        M^E&= f^{(10)}v_d - 3 f^{(126)}w_d + f^{(10')}v'_d - 3 f^{(126')}w'_d
                &(5.3d)\cr}$$
and assume the simple textures
$$\begin{array}{rlrl}
        M^U&= diag(F',\ E',\ C')& \qquad M^{N_{Dirac}}&= diag(-3F',\ -3E',\ C')
		\nonumber\\[0.1cm]
        M^D&= \left(\matrix{0 & A & D\cr A & E & B\cr D & B & C\cr}\right)
		& \qquad
        M^E&= \left(\matrix{F & 0 & D\cr 0 & -3E & B\cr D & B & C\cr}\right)
		\end{array}\eqno(5.4a)$$
with only $D$ complex and the following relations holding
$$\begin{array}{rlrl}
        C'/C&= v_u/v_d,&\qquad E'/E&= w_u/w_d\nonumber\\
        f^{(10')}_{11}v'_d&= -f^{(126)}_{11}w_d = {1\over{4}}F,&\qquad
        f^{(126)}_{11}w_u&= F'\cr
        f^{(10')}_{12}v'_d&= 3f^{(126')}_{11}w'_d = {3\over{4}}A&&\cr
                \end{array}\eqno(5.4b)$$
from which we obtain the constraint, $4F'/F = -E'/E$.

With
$$F' = -\bar{m}_u,\qquad E' = \bar{m}_c,\qquad C' = \bar{m}_t \eqno(5.5a)$$
$$\begin{array}{rlrl}
        C&= 2.4607,& \qquad &{\rm so}\quad v_u/v_d = \tan \beta = 48.9
                \nonumber\\
        E&= -0.3830 \times 10^{-1},& \qquad &{\rm hence}\quad w_u/w_d =
-8.20\cr
        F& = -0.5357 \times 10^{-3},& \qquad B&= 0.8500 \times 10^{-1}\cr
        A& = -0.9700 \times 10^{-2},& \qquad D&= (0.4200 + 0.4285i) \times
                10^{-2}\cr \end{array}\eqno(5.5b)$$
the masses and mixing matrices are calculated at the GUT scale by use of
the projection operator technique of Jarlskog$^{24}$ and then evolved to the
low scales.  The following low-scale results emerge for the quarks:
$$\begin{array}{rlrl}
        m_u(1 {\rm GeV})&= 5.10\ {\rm MeV},& \qquad m_d(1 {\rm GeV})&= 9.33
                \ {\rm MeV}\nonumber\\
        m_c(m_c)&= 1.27\ {\rm GeV},& \qquad m_s(1 {\rm GeV})&= 181\ {\rm MeV}
                \cr
        m_t(m_t)&= 150\ {\rm GeV},& \qquad m_b(m_b)&= 4.09\ {\rm GeV}\cr
  \end{array}\eqno(5.6a)$$
$$V_{CKM} = \left(\matrix{0.9753 & 0.2210 & (0.2089 -0.2242i)\times 10^{-2}\cr
                -0.2209 & 0.9747 & 0.0444\cr
                0.0078 -0.0022i & -0.0438 -0.0005i & 0.9994\cr}\right)
        \eqno(5.6b)$$
These results are in excellent agreement with the input in (2.1a,b), aside from
the unknown CP phase, with $|V_{ub}/V_{cb}| = 0.069$ and
$m_s/m_d = 19.4$, cf. Refs. 11, 12 and 15.

In the absence of any $\bf{126}$ VEV coupling the left-handed neutrino
fields together, we observe that the heavy righthanded Majorana neutrino mass
matrix can be computed at the GUT scale from the approximate seesaw mass
formula$^{25}$
$$M^R = - M^{N_{Dirac}}(M^{N_{eff}})^{-1}M^{N_{Dirac}}\eqno(5.7)$$
$$ = \left(\matrix{(.1744 - .0044i)\times 10^{10} & (-.2332 + .0153i)\times
	10^{11} & (-.2811 - .1925i)\times 10^{12}\cr
	(-.2332 + .0153i)\times 10^{11} & (.6773 - .0329i)\times 10^{12} &
	(-.1189 + .0243i)\times 10^{14}\cr
	(-.2811 - .1925i)\times 10^{12} & (-.1189 + .0243i)\times 10^{14} &
	(.6045 + .0624i)\times 10^{15}\cr}\right)$$
by making use of Eqs. (5.4a) and (5.5a) for the Dirac neutrino matrix and
(4.5d) for the effective light neutrino matix.  Numerically this can be well
approximated by the nearly geometric form
$$M^R = \left(\matrix{F'' & - {2\over{3}}\sqrt{F''E''} &
                -{1\over{3}}\sqrt{F''C''}e^{i\phi_{D''}}\cr
                - {2\over{3}}\sqrt{F''E''} & E'' &
                        -{2\over{3}}\sqrt{E''C''}e^{i\phi_{B''}}\cr
                -{1\over{3}}\sqrt{F''C''}e^{i\phi_{D''}} &
                -{2\over{3}}\sqrt{E''C''}e^{i\phi_{B''}} & C''\cr}\right)
                        \eqno(5.8a)$$
where $E'' = {2\over{3}}\sqrt{F''C''}$ and $\phi_{B''} = - \phi_{D''}/3$.
With $C'' = 0.6077 \times 10^{15},\ F'' = 0.1745\times 10^{10}$ and
$\phi_{D''} = 45^o$, we find numerically
$$\begin{array}{rl}
	M^R = \left(\matrix{0.1745 \times 10^{10} & -0.2307 \times
	10^{11} & (-.2427 - .2427i)\times 10^{12}\cr
	-0.2307\times 10^{11} & 0.6865\times 10^{12} &
	(-.1315 + .0352i)\times 10^{14}\cr
	(-.2427 - .2427i)\times 10^{12} & (-.1315 + .0352i)\times 10^{14} &
	0.6077\times 10^{15}\cr}\right)\end{array}\eqno(5.8b)$$
The structure in (5.8a) can be separated into two parts with
coefficients $2/3$ and $1/3$ which suggests they may arise again from two
different ${\bf 126}$ contributions.  Such geometric textures have been
studied at some length by Lemke$^{26}$ and provide a new mechanism for
leptogenesis as suggested by Murayama.$^{27}$  The resulting heavy Majorana
neutrino masses are found from (5.7b) to be
$$\begin{array}{rl}
M_{R_1}&= 0.249 \times 10^9\ {\rm GeV} \nonumber\\
M_{R_2}&= 0.451\times 10^{12}\ {\rm GeV} \cr
M_{R_3}&= 0.608\times 10^{15}\ {\rm GeV} \cr \end{array}\eqno(5.8c)$$

By making use of the simplified
matrices at the GUT scale first to compute the lepton masses and mixing matrix
$V_{LEPT}$ again by the projection operator technique of Jarlskog$^{24}$ and
then to evolve the results to the low scales, we find at the low scales
$$\begin{array}{rlrl}
        m_{\nu_e}&= 0.534 \times 10^{-5}\ {\rm eV},& \qquad m_e&= 0.504\ {\rm
                MeV}\nonumber\\
        m_{\nu_{\mu}}&= 0.181 \times 10^{-2}\ {\rm eV},& \qquad m_{\mu}&=
                105.2\ {\rm MeV}\cr
        m_{\nu_{\tau}}&= 0.135\ {\rm eV},& \qquad m_{\tau}&= 1.777\ {\rm
                GeV}\cr \end{array}\eqno(5.9a)$$
and
$$V_{LEPT} = \left(\matrix{0.9990 & 0.0451 & (-0.029 -0.227i) \times 10^{-2}
		\cr -0.0422 & 0.9361 & 0.3803\cr
                0.0174 -0.0024i & -0.3799 -0.0001i & 0.9371\cr}\right)
                \eqno(5.9b)$$

The agreement with our starting input is remarkably good, aside from the
CP-violating phases, especially since
only 12 model parameters have been introduced in order to explain 15 masses
and 8 effective mixing parameters.  Although we need two ${\bf 10}$ and
two ${\bf 126}$ Higgs representations for the up, down, charged lepton and
Dirac neutrino matrices with one or two additional ${\bf 126}$'s for the
Majorana matrix, pairs of irreducible representations more naturally emerge
in the superstring framework than do single Higgs representations.  We
have
thus demonstrated by the model constructed that all quark and lepton mass and
mixing data (as assumed herein) can be well understood in the framework of a
simple SUSY GUT model based on SO(10) symmetry.

\subsection{Neutrino Scenario (B) with a 7 eV Tau-Neutrino}

We now apply the same type of reasoning as above to construct an SO(10) model
incorporating a 7 eV tau-neutrino with small mixing with the muon-neutrino
as suggested in the cocktail model$^6$ interpretation of mixed dark matter.  As
noted earlier in Sect. IV, we shall pursue the analysis with the choice of
$x_q = 0$ and $x_{\ell} = 0.3$ which leads to four texture zeros in the quark
mass matrices rather than the simpler SO(10) construct with $x_q = 0.5$ and
$x_{\ell} = 0$, leading to only one texture zero.

Analysis of the down and charged lepton mass matrices in (4.5b) and (4.6a)
with the help of (4.4b) reveals that the observed structure for $M^D$ and
$M^E$ is now approximately given by
$$M^D \sim M^E \sim \left(\matrix{10,126 & 10,126 & 10\cr 10,126 & 126 &
	10,126\cr 10 & 10 & 10\cr}\right)\eqno(5.10)$$
with $M^D_{11}$ again anomalously small.  In fact, since $x_q = 0$ is the
same in both scenarios, the quark mass matrix textures are just $M^U$ and
$M^D$ in (5.4) with the same choice of parameters as listed in (5.5a,b),
since we must fit the same low scale quark masses and CKM mixing matrix.
Only $M^E$ and $M^{N_{Dirac}}$ differ and are modified as given below.
$$\begin{array}{rlrl}
        M^U&= diag(F',\ E',\ C')& \qquad M^{N_{Dirac}}&= diag(-2.5 F',\ -3E',
		\ C')\nonumber\\[0.1cm]
        M^D&= \left(\matrix{0 & A & D\cr A & E & B\cr D & B & C\cr}\right)
		& \qquad
        M^E&= \left(\matrix{{5\over{6}}F & -{1\over{3}}A & D\cr
		-{1\over{3}}A & -3E & {1\over{3}}B\cr
		D & {1\over{3}}B & C\cr}\right) \end{array}\eqno(5.11)$$
with only $D$ again complex.

The heavy righthanded Majorana neutrino mass matrix can again be computed
at the GUT scale from the approximate seesaw mass formula (5.7), and we
find with the help of $M^{N_{Dirac}}$ in (5.11) and $M^{N_{eff}}$ in
(4.6b)
$$M^R = \left(\matrix{(-.1096 + .0059i)\times 10^{11} & (.5523 - .0151i)\times
	10^{11} & (.4594 + .1745i)\times 10^{12}\cr
	(.5523 - .0151i)\times 10^{11} & (.6315 + .0026i)\times 10^{11} &
	(-.2478 - .0939i)\times 10^{13}\cr
	(.4594 + .1745i)\times 10^{12} & (-.2478 + .0939i)\times 10^{13} &
	(-.1733 - .1547i)\times 10^{14}\cr}\right) \eqno(5.12)$$
Numerically this can again be well approximated by the nearly geometric form
$$M^R = \left(\matrix{F'' & - 2 \sqrt{F''E''} &
                - \sqrt{F''C''}e^{i\phi_{D''}}\cr
                - 2 \sqrt{F''E''} & - E'' &
                        2 \sqrt{E''C''}e^{i\phi_{B''}}\cr
                - \sqrt{F''C''}e^{i\phi_{D''}} &
                2 \sqrt{E''C''}e^{i\phi_{B''}} & C''\cr}\right)
                        \eqno(5.13a)$$
where $E'' = {1\over{8}}\sqrt{F''C''}$, aside from an overall sign.
With $C'' = 0.2323 \times 10^{14},\ F'' = 0.1096\times 10^{11}$ and
$\phi_{D''} = 18.7^o$, $\phi_{B''} = 23.0^o$ and $\phi_{C''} = 41.8^o$, we find
numerically
$$\begin{array}{rl}
	M^R = \left(\matrix{-0.1096 \times 10^{11} & 0.5258 \times
	10^{11} & (.4779 + .1618i)\times 10^{12}\cr
	0.5258\times 10^{11} & 0.6307\times 10^{11} &
	(-.2228 + .0946i)\times 10^{13}\cr
	(.4779 + .1618i)\times 10^{12} & (-.2228 - .0946i)\times 10^{13} &
	(-.1732 - .1548i)\times 10^{14}\cr}\right)\end{array}\eqno(5.13b)$$
By means of a phase transformation, one can reduce the three phases to two
without changing the physical content of the mass and mixing matrices.$^{23}$
Again the structure in (5.13a) can be separated into two parts now with
equal coefficients which suggests they may arise from two
different ${\bf 126}$ contributions.
The resulting heavy Majorana neutrino masses for this case are found from
(5.13b) to be
$$\begin{array}{rl}
M_{R_1}&= 0.841 \times 10^9\ {\rm GeV} \nonumber\\
M_{R_2}&= 0.312\times 10^{12}\ {\rm GeV} \cr
M_{R_3}&= 0.235\times 10^{14}\ {\rm GeV} \cr \end{array}\eqno(5.13c)$$

By again making use of the simplified
matrices at the GUT scale first to compute the lepton masses and mixing matrix
$V_{LEPT}$ by the projection operator technique of Jarlskog$^{24}$ and then
to evolve the results to the low scales, we find at the low scales for the
(B) scenario
$$\begin{array}{rlrl}
        m_{\nu_e}&= 0.544 \times 10^{-6}\ {\rm eV},& \qquad m_e&= 0.511\ {\rm
                MeV}\nonumber\\
        m_{\nu_{\mu}}&= 0.242 \times 10^{-2}\ {\rm eV},& \qquad m_{\mu}&=
                107.9\ {\rm MeV}\cr
        m_{\nu_{\tau}}&= 6.99\ {\rm eV},& \qquad m_{\tau}&= 1.776\ {\rm
                GeV}\cr \end{array}\eqno(5.14a)$$
and
$$V_{LEPT} = \left(\matrix{0.9992 & 0.0410 & (0.150 -0.107i) \times 10^{-2}
		\cr -0.0411 & 0.9991 & 0.0113\cr
                -0.0010 -0.0011i & -0.0123 & 0.9999\cr}\right)
                \eqno(5.14b)$$

\section{SUMMARY}

In this paper we have demonstrated how one can apply the procedure outlined
in the introduction to two different neutrino scenarios to construct model
mass matrices which fit well the low energy input data assumed at the
outset.  The full set of quark, lepton and neutrino masses and their mixing
matrices are evolved to the grand unification scale, where the mass matrices
can be constructed numerically by an extension of Kusenko's method which he
applied only
to quarks.  A key ingredient is the possibility to vary two parameters,
$x_q$ and $x_l$, as well as the signs of the masses in the diagonal matrices,
which allow one to scan the bases for a choice where the mass matrices may
exhibit simple SO(10) structure.  Knowledge of the preferred
bases and SO(10) symmetry structure of the up, down, charged lepton and Dirac
neutrino mass matrices then allows one to construct a set of model matrices
with a small set of parameters.  From the numerical light neutrino
and Dirac neutrino mass matrices, one can then deduce the structure of the
heavy right-handed Majorana mass matrix.

In scenario (A) where one makes use of data on the nonadiabatic MSW
interpretation of the solar neutrino flux depletion as well as the observed
depletion of atmospheric muon-neutrinos, we identified a simple
SO(10) structure for the selection of $x_q = 0$ and $x_{\ell} = 0.88$.
In this basis, the
up quark matrix is real and diagonal, while the down quark, charged lepton
and light neutrino mass matrices are complex symmetric.  Not only is the
symmetry structure simple, but the maximum number of texture zeros is obtained
for the mass matrices.  In particular, we find that a minimum of two
${\bf 10}$'s and two ${\bf 126}$'s of Higgs representations are required for
the mass generation, while only four texture zeros appear in the up and down
quark mass matrices.  This should be contrasted with other authors'
assumptions$^{4,5}$ of just one set of ${\bf 10}$'s and ${\bf 126}$'s or a
minimum of five texture zeros.

In scenario (B), data on the nonadiabatic MSW solar depletion effect is used
together with that for a 7 eV tau-neutrino which provides the hot dark matter
component of mixed dark matter.  In this case the choice of $x_q = 0.5$ and
$x_{\ell} = 0$ provides the simplest SO(10) structure, but only one texture
zero appears.  As an alternative, we have sacrificed some structure simplicity
to maintain four texture zeros in the up and down quark matrices by selecting
instead $x_q = 0$ and $x_{\ell} = 0.3$.  The up and down quark mass matrices
then have exactly the same structure as in scenario (A), for the same quark
masses and CKM mixing matrix must be obtained as before, but the charged
lepton and neutrino matrices are now different.  In view of the goal of
constructing quark and lepton mass matrices which exhibit the simplest SO(10)
structure and the largest number of texture zeros, we conclude that
scenario~(A) is favored over that for scenario~(B).

We have also explored the sensitivity of our results to some changes in the
input parameters.  In particular, we find that the results obtained change
little if one varies the electron-neutrino mass from $10^{-4}$ eV to
$10^{-10}$ eV.  The only pronounced effect is a change in the heavy
right-handed Majorana mass matrix and its eigenvalues which can shift by
one order of magnitude for the range of $m_{\nu_e}$ considered.  Likewise,
we find that if one lowers $V_{cb}$ to 0.038, the simple SO(10) structure
is retained in scenario (A), for example, with $x_q = 0$ and $x_{\ell} =
0.90$; similarly for scenario (B).  Small changes in the model parameters
occurring in (5.5) then again permit good reproduction of the initial input
data.

Work is now underway to try to identify discrete symmetries or family
symmetries which will lead to the matrix models constructed at the GUT
scale.  We also are studying a similar type of analysis for higher
symmetry groups such as SO(18).
\newpage

The authors gratefully acknowledge the Summer Visitor Program and warm
hospitality of the Fermilab Theoretical Physics Department, which enabled the
initiation of this research.  We thank Joseph Lykken for his continued interest
and comments during the course of this work.  The research of CHA was
supported in part by Grant No. PHY-9207696 from the National Science
Foundation, while that of SN was supported in part by the U.S. Department of
Energy, Grant No. DE-FG05-85ER 40215.

\begin{reflist}
\item	C. H. Albright and S. Nandi, Fermilab-PUB-93/316-T and OSU Preprint
	282, submitted for publication.

\item   H. Fritzsch, Phys. Lett. {\bf 73B}, 317 (1978); Nucl. Phys. B
	{\bf 155}, 182 (1979).

\item   H. Georgi and C. Jarlskog, Phys. Lett. {\bf 86B}, 297 (1979);
 	J. A. Harvey, P. Ramond and D. B. Reiss, Phys. Lett. {\bf 92B}, 309
	(1980); H. Arason, D. Castan\~{n}o, B. Keszthelyi, S. Mikaelian,
	E. Piard, P. Ramond and B. Wright, Phys. Rev. Lett. {\bf 67}, 2933
	(1991); Phys. Rev. D {\bf 46}, 3945 (1992); H. Arason, D. Castan\~{n}o,
	P. Ramond and E. Piard, Phys. Rev. D {\bf 47}, 232 (1993);
	A. Kusenko and R. Shrock, Stony Brook preprint ITP-SB-93-37, to be
	published.

\item 	S. Dimopoulos, L. J. Hall and S. Raby, Phys. Rev. Lett.
	{\bf 68}, 1984 (1992);  Phys. Rev. D {\bf 45}, 4192 (1992);
	{\bf 46}, R4793 (1992); {\bf 47}, R3702 (1993);  G. F. Giudice,
	Mod. Phys. Lett. A {\bf 7}, 2429 (1992); H. Arason, D. J.
	Casta\~{n}o, E.-J. Piard and P. Ramond, Phys. Rev. D {\bf 47},
	232 (1993); P. Ramond, R. G. Roberts and G. G. Ross, Nucl.
	Phys. {\bf B406}, 19 (1993); V. Barger, M. S. Berger, T. Han and M.
	Zralek, Phys. Rev. Lett. {\bf 68}, 3394 (1992); Phys. Rev. D {\bf 47},
	2038 (1993); G. W. Anderson, S. Raby, S. Dimopoulos, L. J. Hall and G.
	D. Starkman, Lawrence Berkeley Laboratory preprint LBL-33531.

\item   K. S. Babu and R. N. Mohapatra, Phys. Rev. Lett. {\bf 70},
        2845 (1993); L. Lavoura, Phys. Rev. D {\bf 48}, 5440 (1993).

\item   Q. Shafi and F. Stecker, Phys. Rev. Lett. {\bf 53}, 1292 (1984);
        Ap. J. {\bf 347}, 575 (1989); G. G. Raffelt, Max Planck Institute
        preprint, MPI-Ph/93-81; J. Ellis and P. Sikivie, CERN preprint,
        CERN-TH.6992/93.

\item   K. S. Hirata et al., Phys. Lett. B {\bf 205},416 (1988); D. Casper
                et al., Phys. Rev. Lett. {\bf 66}, 2561 (1991);

\item   A. Kusenko, Phys. Lett. B {\bf 284}, 390 (1992).

\item   S. P. Mikheyev and A. Yu Smirnov, Yad Fiz. {\bf 42}, 1441 (1985)
                [Sov. J. Nucl. Phys. {\bf 42}, 913 (1986)]; Zh. Eksp. Teor.
                Fiz. {\bf 91}, 7 (1986) [Sov. Phys. JETP {\bf 64}, 4
                (1986)]; Nuovo Cimento {\bf 9C}, 17 (1986); L. Wolfenstein,
                Phys. Rev. D {\bf 17}, 2369 (1978); {\bf 20}, 2634 (1979).

\item	R. Davis et al., Phys. Rev. Lett. {\bf 20}, 1205 (1968); in {\it
	Neutrino '88}, ed. J. Schnepp et al. (World Scientific, 1988);
	K. Hirata et al., Phys. Rev. Lett. {\bf 65}, 1297, 1301 (1990);
	A. I. Abazov et al., Phys. Rev. Lett. {\bf 67}, 3332 (1991);
	Phys. Rev. C {\bf 45}, 2450 (1992); P. Anselmann et al., Phys. Lett.
	B {\bf 285}, 376, 390 (1992).

\item	J. F. Donoghue, B. R. Holstein and D. Wyler, Phys. Rev. Lett. {\bf 69},
	3444 (1992).

\item   J. Gasser and H. Leutwyler, Phys. Rep. C {\bf 87}, 77 (1982).

\item	F. Abe et al., Fermilab preprint, FERMILAB-CONF-93-212-E; S. Abachi
	et al., Fermilab preprint FERMILAB-PUB-94-004-E.

\item	N. Cabibbo, Phys. Rev. Lett. {\bf 10}, 531 (1963); M. Kobayashi
	and T. Maskawa, Prog. Theor. Phys. {\bf 49}, 652 (1973).

\item   Review of Particle Properties, Phys. Rev. D {\bf 45}, No. 11,
        Part II, June (1992); cf. also
	S. Stone, to appear in the Proceedings of the Yukawa Workshop held
	at the University of Florida, Feb. 11-13, 1994.

\item   Two recent studies of the atmospheric neutrino flux issues are
	reported in T. K. Gaisser, Bartol Res. Inst. preprint BA-93-62; D. H.
	Perkins, Oxford preprint OUNP-93-32.

\item   N. Ushida et al., Phys. Rev. Lett. {\bf 57}, 2897 (1986).

\item   See, for example, C. H. Albright, Phys. Rev. D {\bf 45}, R725 (1992);
	Zeit. f. Phys. C {\bf 56}, 577 (1992); A. Yu. Smirnov, Phys. Rev.
	D {\bf 48}, 3264 (1993).

\item   S. Naculich, Phys. Rev. D {\bf 48}, 5293 (1993).  The complete two
	loop calculations can be found in V. Barger, M. S. Berger and
	P. Ohmann, Phys. Rev. D {\bf 47}, 1093 (1993); G. W. Anderson,
	S. Raby, S. Dimopoulos and L. J. Hall, Phys. Rev. D {\bf 47},
	R3702 (1993).

\item	B. Ananthanarayan, G. Lazarides and Q. Shafi, Phys. Rev. D {\bf 44},
	1613 (1991); cf. also the last two citations in Ref. 19.

\item	The two loop RGEs would lead to somewhat different $\tan \beta$ and
	$m_b(m_b)$.

\item   F. R. Gantmakher, Teoriya Matrits (Nauka, Moscow, 1966).

\item	A general analysis of the number of phases required in the various
	quark and lepton mass matrices has been carried out in
	A. Kusenko and R. Shrock, Stony Brook preprints ITP-SB-93-58,62,63;
        cf. also A. Santamaria, Phys. Lett. B {\bf 305}, 90 (1993).

\item   C. Jarlskog, Phys. Rev. D {\bf 35}, 1685 (1987); {\bf 36}, 2138
                (1987); C. Jarlskog and A. Kleppe, Nucl. Phys. {\bf B286},
                245 (1987).

\item   M. Gell-Mann, P. Ramond, and R. Slansky, in {\it Supersymmetry},
                edited by P. Van Nieuwenhuizen and D. Z. Freedman
                (North-Holland, Amsterdam, 1979); T. Yanagida, Prog. Theor.
                Phys. {\bf B 315}, 66 (1978).

\item   E. H. Lemke, Mod. Phys. Lett. A {\bf 7}, 1175 (1992).

\item	H. Murayama, to appear in the Proceedings of the Yukawa Workshop
	held at the University of Florida, Feb. 11-13, 1994.
\end{reflist}
\end{document}